%% file: arxiv.tex
\documentclass[11pt]{article}

\usepackage{fullpage}
\usepackage{amsmath}
\usepackage{amssymb}
\usepackage{subfigure}
\usepackage{graphicx}
\usepackage{anysize}
\marginsize{2.5cm}{2cm}{2cm}{2.5cm}

\usepackage[ruled]{algorithm2e}

\SetAlFnt{\small}
\SetAlCapFnt{\small}
\SetAlCapNameFnt{\small}
\SetAlCapHSkip{0pt}
\IncMargin{-\parindent}
\input{math.tex}
\long\def\comments#1{}

\newcommand{\Dset}{\mathcal{D}}

\newcommand{\Pset}{\mathcal{P}}

\renewcommand{\and}{\qquad}

\begin{document}


\title{Refining Recency Search Results with User Click Feedback}
\author{Taesup Moon \and Wei Chu \and Lihong Li \and Zhaohui Zheng \and Yi Chang \\
\{taesup,chuwei,lihong,zhaohui,yichang\}@yahoo-inc.com \\
Yahoo! Labs, 701 First Avenue, Sunnyvale, CA, USA 94089}
\date{}
\maketitle

\begin{abstract}
Traditional machine-learned ranking systems for web search are
often trained to capture stationary relevance of documents to
queries, which has limited ability to track non-stationary user
intention in a timely manner. In recency search, for instance, the relevance of
documents to a query on breaking news often changes significantly over time, requiring effective adaptation to user intention. In this
paper, we focus on recency search and study a number
of algorithms to improve ranking results by leveraging
user click feedback.  Our contributions are three-fold.  First, we use
real search sessions collected in a random exploration bucket for
\emph{reliable} offline evaluation of these algorithms, which provides
an unbiased comparison across algorithms without online bucket tests.
Second, we propose a re-ranking
approach to improve search results for recency queries using user clicks.
Third, our empirical comparison of a dozen algorithms on real-life search data suggests
importance of a few algorithmic choices in these applications,
including generalization across different query-document pairs,
specialization to popular queries, and real-time adaptation of
user clicks.
\end{abstract}






\section{Introduction}
\label{sec: introduction}

Ranking a list of documents based on their relevance to a given
query is the central problem in various search applications of the
Internet.
%
%
Machine-learned ranking algorithms have been shown highly
effective for generalizing to unseen data from labeled training
data and have been very successful especially in commercial Web
search engines; see~\cite{joachims2002,BurLeRag07,BurShaRenLazetal05,FreIyeSchSin03,icml07cortes,ZheZhaZhaChaChe08,Liu09} and many references therein for more reference. Usually, such machine-learned ranking algorithms
learn a ranking function based on editorial judgments---relevance
labels provided by human editors.
A critical assumption here is that the relevance of documents for
a given query is more or less stationary over time, and therefore,
as long as the coverage of training set is broad enough, the
ranking function learned from the training set would be sufficient
to generalize to unseen data in the future.  This assumption is
often valid in web search, especially for popular queries like
``\emph{yahoo}'', for which document relevance is indeed (almost)
static.

However, there are other important categories of applications
where document relevance to a query may change over time.  One
such example is the recency ranking problem in web search: when
breaking news emerges, a document that used to be most relevant to
a
query may be superseded by others that have more relevant
information about the news; see Section~\ref{sec: motivation} for
a concrete example.
A key challenge for such problems is to track user intention in a
timely fashion.

An interesting attempt was taken recently for tracking
non-stationary document relevance~\cite{Anlei10}. The authors
devised time-varying features that reflect freshness of documents
and utilized recency demoted labels provided by human editors that
explicitly modify the relevance target values
in the training set. Their results showed an improvement of
ranking qualities for time-sensitive queries.
However, their approach is still based on editorial judgments and
so limited
for two reasons. First, obtaining high-quality training data is
hard.  Implementing more fine-grained time-varying features, such
as features from the time series of clicks that can accurately
follow the relevance drifts is considerably subtle and complex
since carefully testing and selecting good features is a long and
complicated process.
Also, obtaining laborious recency demoted labels from human
editors not only is expensive, but also can be inaccurate in
correctly representing the temporal variation of document
relevance. Second, even when we can come up with such complex and
expensive data to batch-train a ranking function, tracking actual
user intention remains challenging due to the very unpredictable
nature of how user intention evolves over time.

In this paper, we investigate how to leverage user click feedback
to complement and improve such editorial-judgment based ranking
systems.
Our rationale is that, particularly for recency queries, instantaneous click trends on the top portion of the ranking list 
are important indicators of document relevance. Such signals
allow us to extract subtle information that may be hard for
human editors to foresee when they provide relevance judgments.
In particular, we explicitly track the click-through rate (CTR) of
a query-document pair using a linear combination of extracted
features, including the editorial-judgment based ranking function's score. Based on search results returned by the current
search engine, we propose a re-ranking approach to further improve
search results for recency queries. We use user click as labels
for training the CTR models in either batch or online mode.

In order to reliably evaluate and compare our algorithms, an
``exploration bucket'' was set up for a small random portion of
live traffic for recency-classified queries in a commercial search
engine.  Within the bucket, the top URLs returned by the search
engine was randomly shuffled.
This bucket is critical to the work reported in this paper for two
reasons. First, it provides a mechanism for \emph{exploration}
that is essential for interactive learning problems as the one
considered in this paper to get rid of evaluation bias (Section~\ref{subsec: bucket}).
Second, it allows us to
obtain \emph{unbiased} evaluation of algorithms without the need
for online bucket tests (Section~\ref{subsec: evaluation method}).

This work extensively augments our preliminary results reported in an extended abstract~\cite{MooLiChuLiaZheCha10}, and 
uses random exploration data to do \emph{unbiased}
evaluation on a dozen of algorithms that improve ranking by leveraging user
click feedback on a major web search engine.

The rest of the paper is organized as follows.  We describe our
exploration bucket in Section~\ref{subsec: bucket}.  Using data in this bucket,
we present a
motivating example in Section~\ref{sec: motivation}, showing the necessity of taking into account temporal variation of document relevance reflected in user click feedback.  Our methods
are detailed in Section~\ref{sec: our method} and empirically
evaluated using the exploration bucket data in Section~\ref{sec:
experimental results}.  We then discuss related work in
Section~\ref{sec: related work} and conclude the paper in
Section~\ref{sec: conclusion}.

\section{Exploration Bucket Data}\label{subsec: bucket}

As described in the previous section, we set up a bucket to collect
exploration data from a small portion of live traffic from a
commercial search engine. The bucket started on Jan 29, 2010
and ended on Feb 4, 2010. Throughout these days, we collected $399,880$ search
sessions that contained 61,904 recency classified queries,
after removing non-random sessions corrupted by business rules.
%
The ranked list for those queries were generated by the recency
ranking function trained as described in \cite{Anlei10} and the
ranking score for each query-document was recorded. For each
session, we randomly shuffled the top four results and logged the
permutation id of each shuffled permutation (a total of $4!=24$ of
them) and user clicks on the corresponding permuted ranking
results.

\begin{figure}[ht]
\centering
\includegraphics[width=0.5\textwidth]{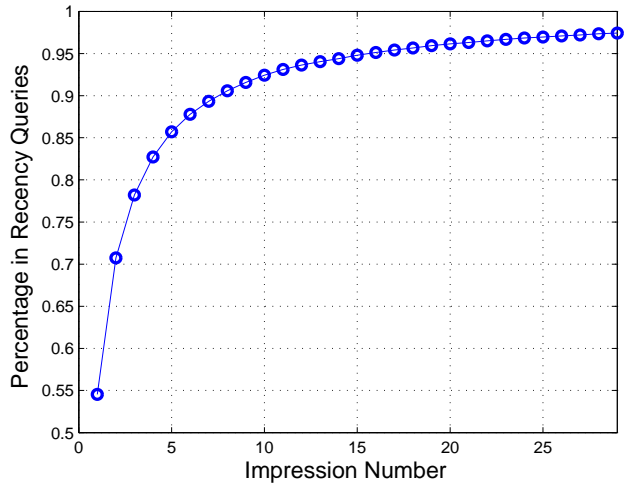}
\caption{Recency query impression.}\label{fig:impression}
\end{figure}

The collected data is very sparse and long-tailed as shown in 
Figure~\ref{fig:impression}, in which $92.4\%$ of queries were
issued no more than $10$ times and more than half of queries were
issued just once.  The reason for this sparsity is that the
recency query classifier utilizes some language model to determine
the queries that are related to each other, which causes some
recency-related idiosyncratic, less popular queries---such as
different word orderings or typographically wrong queries---to
be classified as recency queries.

By doing the random shuffling, we are able to collect user click
feedback on each document without positional bias, and such
feedback can be thought of as a reliable proxy on relevance of
documents. Note that the effect on user experience of shuffling
would not be as severe as that for navigational queries, since the
relevance differences of top-ranked documents to recency queries would not be
as dramatic as those for navigational queries. Also, we chose a
reasonably small number, $4$, in order to limit the negative impact on user
experience in the exploration bucket.

\begin{figure}[t]
\centering
\includegraphics[width=0.7\textwidth]{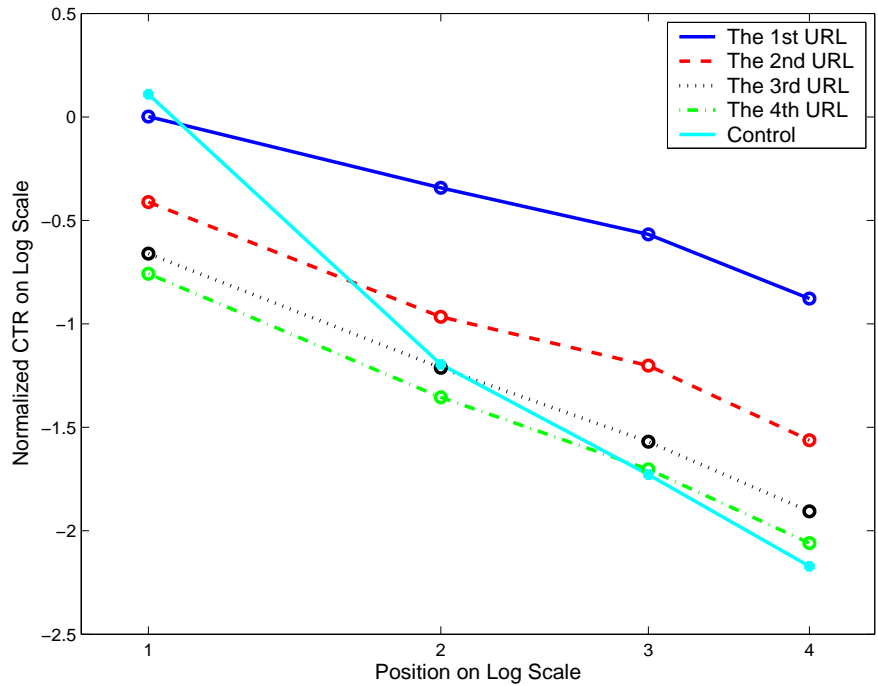}
\caption{Marginal nCTRs of four types of URL at the top four
positions in exploration bucket, along with the nCTRs on the
original order as control. All are on logarithm
scale.}\label{fig:positionallogctr}
\end{figure}

Another byproduct of our exploration bucket is that we can accurately observe the positional biases of clicks.
Specifically, for the top four URLs in each session, we
can infer the original ranking of the search engine simply by
the recorded ranking scores.  For each session, the URL with a highest
ranking score is called the ``1st URL''.  These URLs were displayed
an equal number of times in all four positions in the exploration bucket
because of the random shuffling.  We can then estimate the aggregate
Click-through rate (CTR) of the 1st URLs in each of the four positions,
as depicted by the blue line in Figure~\ref{fig:positionallogctr}.\footnote{To
protect business-sensitive information, the paper only reports the
\emph{normalized CTR} or \emph{nCTR}, which is the CTR multiplied
by a constant.  We will use them interchangeably if there is no confusion.}
Such nCTRs are \emph{marginal} since we have taken all possible layouts
(of other URLs) into account, thanks to the uniform randomness
in the exploration bucket.  Similarly, marginal nCTRs of the 2nd, 3rd, and 4th
URLs of all sessions are also plotted in Figure~\ref{fig:positionallogctr}.

Interestingly, lines of these four marginal nCTRs are almost parallel
to each other, which implies the user click patterns follow the
well-known power-law distribution. The slope indicates the intrinsic
positional biases in the displayed layout of search results.
To further illustrate the conditional effect on user
click patterns, we also present the nCTRs of the original display
order.  It corresponds to the steeper straight line of
Figure~\ref{fig:positionallogctr}, indexed by ``Control'' in cyan.
We observed that the nCTR of the 2nd URL at the 2nd position
conditioned on the 1st URL at the 1st position is much lower than
the marginal CTR of the 2nd URL at the 2nd position. The drop
indicates a negative conditional effect from the 1st URL at the
1st position. For the 3rd URL at the 3rd position and the 4th URL
at the 4th positions, we observed the similar conditional effect. \\

The apparent positional biases shown in Figure~\ref{fig:positionallogctr} would be taken into account below in devising our re-ranking algorithm.
Moreover, the fact that the four lines other than the ``Control'' do not cross each other shows that, \emph{on average}, the original ranking is doing a decent job in ranking the URLs also with respect to CTRs. However, in this paper, we show that we can do better than this by re-ranking the URLs appropriately
so that the overall CTRs of re-ranked results can be further improved.

\section{Motivation} \label{sec: motivation}

Before running into technical details, let us
first look at a concrete example found in exploration bucket to
illustrate our motivation, and then summarize the challenges that
we confront in recency search results.  This example will be revisited
in our discussion of experimental results.

\begin{figure}[t]
\centering
\includegraphics[width=0.75\textwidth]{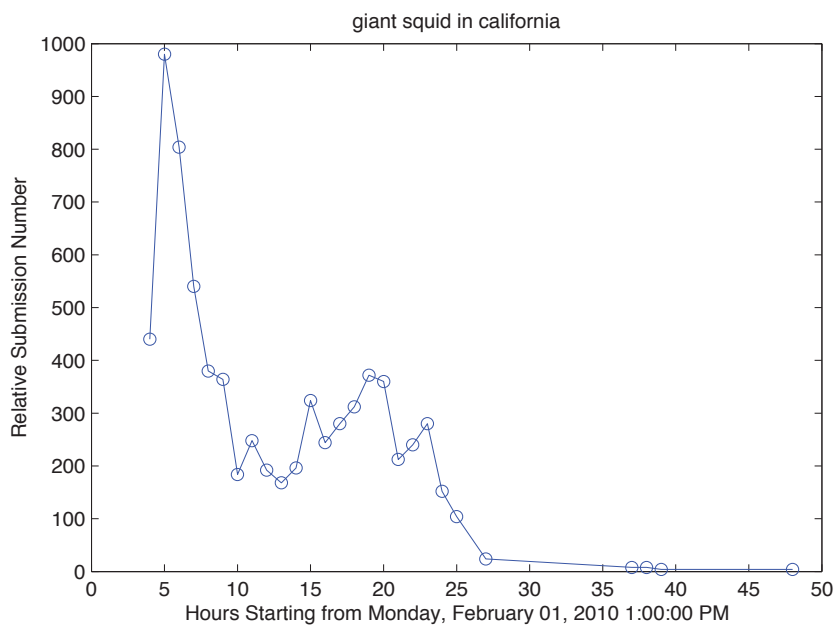}
\caption{Impression count in the exploration bucket of the
query ``\emph{giant squid in California}''.}\label{fig:squid}
\end{figure}

As in \cite[Section 4]{Anlei10}, recency queries are defined to be
time-sensitive queries that show non-stationary temporal
statistics compared to the past query logs. The query,
``\emph{giant squid in California}'', is a typical recency query,
which appeared on February 1, 2010 and then disappeared after two days
in our exploration bucket data. Figure \ref{fig:squid} shows the
impression statistics of the query with respect to
time.\footnote{To avoid revealing business-sensitive data, we
normalize the query submission number by multiplying it with a positive number.}
This query is related to local news in California. At that time,
a number of giant squids weighing up to 60 pounds had swum into
waters off the Californian coast and were caught by sport
fishermen by the hundreds. To find related materials of the
local news, many people submitted the query ``\emph{giant
squid in California}'' to search engines.

\begin{figure}[t]
\centering
\includegraphics[width=0.65\textwidth]{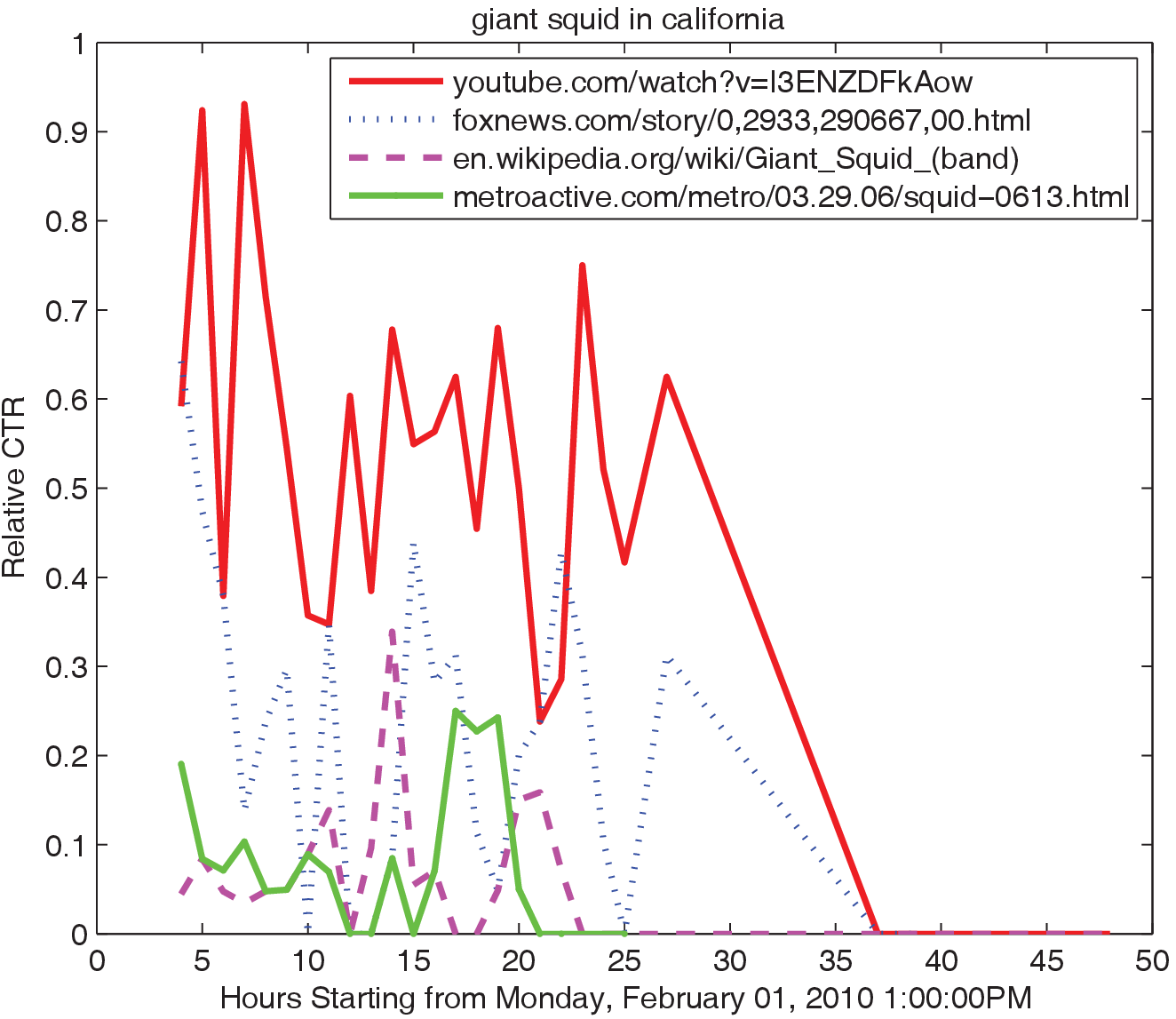}
\caption{Hourly CTR of the top 4 URLs associated with the recency
query ``\emph{giant squid in California}''. After the 30th hour,
no more click of this query was observed in the exploration
bucket.}\label{fig:squidctr}
\end{figure}

To study user click patterns on the URLs associated with the
query, we again examined the top four URLs in the exploration bucket.
The four URLs were retrieved by the default ranking function for
recency queries in the search engine, which was trained on
editorial judgments in batch mode. The ranking generated by the
search engine on the four URLs was:
\begin{enumerate}
\item \verb"foxnews.com/story/0,2933,290667,00.html"
\item \verb"en.wikipedia.org/wiki/Giant_Squid_(band)"
\item \verb"metroactive.com/metro/03.29.06/squid-0613.html"
\item \verb"youtube.com/watch?v=I3ENZDFkAow"
\end{enumerate}
%
%

Based on their contents, the four web pages can be categorized as ``a
news story page'', ``a background knowledge page'', ``a relevant
page'', and ``a video page''. As we randomly shuffled the display
order of the top 4 URLs in the exploration bucket, each URL had
the same chance to be displayed at each position. Therefore,
position bias on clicks for these URLs was removed. The nCTRs
of the four URLs observed in our exploration
bucket session data are presented in Figure \ref{fig:squidctr}.
Clearly, although the initial nCTRs were similar, the
``video'' content ended up receiving most clicks, while the ``news
story'' was runner-up. This shows that while our recency ranking
function made a reasonable decision (by putting the ``video''
content within top 4), it yet failed to accurately predict the
ranking with respect to the users' preference reflected in the CTR
patterns on the URLs. Moreover, we note that unless we actually
see this click patterns, it would be extremely hard for human
editors to predict such relevance patterns.


%

%


Based on the above observations, two challenges are identified for the recency ranking problem.

%


\begin{itemize}

\item \emph{Relevance Drifting}: As 
illustrated by the case above, document relevance may vary significantly over time. The above examples show the limitations of the editorial judgment based, batch learning framework in tracking such temporal dynamics. In general, it is very difficult to design features that can correctly reflect the temporal variances of relevance and for editors to predict the relevance labels before observing the actual user behaviors. How to rapidly track such drift would be a major challenge of recency ranking.


\comments{
\item \emph{Dynamic Content}: The Web is highly dynamic,
with millions of fresh contents being generated and linked
every second. Frequently, we cannot curate informative features
for fresh content in time, especially users' feedback.
How to identify impressive content quickly becomes a critical
issue for recency queries. }

\item \emph{Data Sparsity}:  Due to the reason specified in Section \ref{subsec: bucket}, many recency queries have few impressions.
Hence, learning across queries (\textit{i.e.}, generalization) would be important.


\end{itemize}

%

In addition, we note that keeping track of dynamic content for recency queries to generate reasonably good top documents is also a critical challenge, but, it is outside the scope of the paper. In the present work, we rely on the baseline search engine to retrieve the most relevant documents, and focus on improved re-ranking using click feedback.

\section{Our method}\label{sec: our method}

To address the two challenges in the previous subsection,
we believe that it is helpful for a ranking module to detect and track the
non-stationarity of user interests reflected on the click patterns. The example in Section \ref{sec: motivation} suggests that, when positional biases are removed, CTRs can be important indicators of relevances of URLs, especially for recency queries. Moreover, as shown in
Figure~\ref{fig:positionallogctr}, it is also important to remove
positional biases and conditional effects in CTR estimates for
documents. To this end, we assume that CTR at the top position, denoted by CTR@1, is of minimal conditional effect, and thus, we use CTR@1 as a proxy of relevance of a URL for learning and evaluating our method.


Ideally, if we knew the true CTR@1 of
every document for a query,
we could display the document with the highest CTR@1 at the top position.
However, keeping track of CTR@1 of \emph{all} query-document pairs is
unnecessarily difficult: similar to general Web search, it is more
important to get better estimates for \emph{high-quality}
documents for queries. Therefore, assuming that our original ranking function retrieves reasonably good quality documents at the top,  we propose a
\emph{re-ranking approach} that estimates CTR@1 of top-ranked
documents retrieved by the current search engines and then
optimizes the ranking order by CTR@1 estimates. Our approach is
designed specifically to address the challenges mentioned in
Section~\ref{sec: motivation} for recency queries:

\begin{itemize}

\item \emph{Relevance Drifting}: In contrast to waiting for
editorial judgments to update ranking results for a recency query,
our algorithm updates CTR estimates near real-time based on user
click feedback on the ranked list of documents. Not only does it
avoid the expensive editorial judgments, but it can also quickly
adapt to the varying relevance and then maximize CTR.


\comments{
\item \emph{Dynamic Content}:
Our algorithm updates its model parameters based on user click
feedback in exploration traffic, refer to
Section~\ref{subsec:param-update} for more details. Note that we
still rely on the search engine to retrieve the candidates of
fresh high-quality documents, but our approach can actively
collect click feedback on potentially relevant documents that are
not easily captured by existing features.
%
}

\item \emph{Data Sparsity}: Our algorithm works in a common feature space shared by \emph{all} query--document pairs.  It is then able to generalize click feedback of a pair to other pairs via feature values.
Furthermore, we will also show the benefits of maintaining bias
terms, or latent features, dedicated to popular query--document pairs in the experimental results.

%

\end{itemize}


%

%


%







One may question the appropriateness of using CTR@1 as a target for ranking problems.  However, this choice is justified, particularly for the recency ranking problem, for the following reasons.  First, CTR@1 is already an important metric that is considered for deploying general machine-learned ranking function for web search in practice. Second, our example in Section \ref{sec: motivation} shows that CTR@1 can be a more objective metric than editorial labels for the recency queries, for which relevance judgments are difficult to obtain.  Third, our approach does not completely ignore editorial judgments, because only the top portion of ranking result list is refined.  Since the baseline ranking list is obtained based on editorial labels, traditional relevance-oriented metrics like NDCG (Normalized Discounted Cumulative Gain)~\cite{JarKek02b} are already reasonably high.  Finally, it is worth noting that our approach can also be used to maximize other metrics of interest such as session length or revenue.  With these reasonings, we now describe the setting and algorithm of our approach more in detail.

\subsection{Settings} \label{subsec: notations}
We consider the following re-ranking framework, naturally modeled as a
round-by-round process: at round $t$,
\begin{enumerate}
\item{A user arrives and types in a query $q_t$.}
\item{The default recency ranking function generates an ordered list of $s$ documents with highest relevance scores. Then, our re-ranking function re-orders these $s$ documents and present to the user the re-ordered ranked list $\{u_{t,1},\ldots,u_{t,s}\}$.
}
\item{The user then provides feedback $r_t=\{c_{t,1}, \ldots, c_{t,s}\}$ on our re-ranking result, where $c_{t,i}=1$ if a user clicked on the document at potision $i$, and $0$ otherwise. }
\item{Based on the user feedback $r_t$, the re-ranking function is updated and is used for the next round $t+1$.}
\end{enumerate}
From above described process, we see that our re-ranking function is inherently an online algorithm that updates its logic on the fly from the sequential observation of click feedback. 
In order to efficiently implement and update our re-ranking function, we implement a common feature vector for every query-document pair, $(q,u)$, and denote it as $\mathbf{x}_{qu}\in\mathbb{R}^d$. In our experiment, a total of $d=51$ features were used.  These features include
regular query-specific (e.g., number of words in a query), document-specific (e.g., spam classification score of a document), and
query-document-specific (e.g., number of times a query appears in a given document) features used in ordinary machine learned
ranking functions, and more importantly, the ranking score
generated by the default, editorial judgment-based recency ranking function. 
Our re-ranking function is then defined to be a function that predicts the CTR@1 of each $(q,u)$ as a function of $\mathbf{x}_{qu}$ and possibly of some latent features, and the function gets updated based on observing users' click feedback in an online fashion. The detailed function form and update formula are described in the subsequent two sections.

Given our goal of maximizing CTR@1 in the re-ranking
results, it is tempting for an online algorithm to follow a
\emph{greedy} strategy: that is, it \emph{always} ranks (for the
present query at hand) the documents in the order of the highest CTR@1 estimates and updates the function parameters solely based on the user feedback for the algorithm's ordered list. While this greedy approach is intuitively desirable, it can be detrimental in practice. 
This is because, as can be seen in the interactive round-by-round process described above, the re-ranking algorithm obtains
user feedback \emph{only} from the orderings that it has displayed to the user. Therefore, if an algorithm mistakenly orders the documents, a greedy re-ranking strategy can prevent it from collecting user feedback for other (potentially better) rankings and correcting its mistake to find the most relevant document on the top for maximizing CTR@1.  
Consequently, the algorithm has to balance
two conflicting goals: (a) ``exploitation'' --- to display in the
first position most relevant documents to maximize re-ranking
quality (in our case, to maximize user clicks), and (b)
``exploration'' --- to display documents for the purpose of
collecting data to further improvement.
The exploration/exploitation tradeoff described above is a
defining characteristic of a class of problems known as bandit
problems~\cite{Robbins52Some}, which has received considerable
attention recently for Internet-related
applications~\cite{RadJoa07,YueJoa09}.

In the following Section \ref{subsec:param-ctr} and Section \ref{subsec:param-update}, we present the basic function form of our online re-ranking function and its update formula provided the user feedback $r_t$ is given, respectively. Then, in Section \ref{subsec:variation}, we describe how we vary our scheme in order to cope with the explore/exploit tradeoff explained above and accelerate learning speed. 

\subsection{Parametric CTR@1 Estimate} \label{subsec:param-ctr}

Although many alternatives exist, we choose our re-ranking
function to be linear in the feature vector $\mathbf{x}_{qu}$.
This choice allows us to derive \emph{exact} update rules and simplify the exposition.
Other non-linear models may also be used, although numerical approximation is unavoidable in general when optimizing their model parameter.  In particular, we have tried logistic regression and the probit-based regression~\cite{Graepel10Web}, and observed similar performance as the linear model.

%
%
Since we try to maximize CTR@1, it is natural to find a function
that estimates CTR@1 of a $(q,u)$ pair for re-ranking. Once the
feature vector $\vecb{x}_{qu}$ of length $d$ is given for a
$(q,u)$ pair, a linear combination of them is used to estimate
CTR@1.  In fact, we will use the most general form that captures
all variants useful in our experiments:
\begin{eqnarray}
\text{CTR@1}(q,u) &=& \pmb{\beta}^{\mt}\vecb{x}_{qu} + b_{q,u},
\label{eqn:model-hybrid}
\end{eqnarray}
where the vector $\pmb{\beta}\in\mathbb{R}^{d}$ contains the
coefficients shared by \emph{all} query-document pairs, and
$b_{q,u}\in\mathbb{R}$ is a $(q,u)$-specific \emph{bias} term.  Both
$\pmb{\beta}$ and $\{b_{q,u}\}$ are to be learned by our
algorithm.




%

Clearly, user click feedback on any query-document pair may be
used to estimate $\pmb{\beta}$, which in turn can be used to
predict CTR@1 for other query-document pairs.  Therefore, the
linear part of $\pmb{\beta}$ in Equation~\ref{eqn:model-hybrid}
addresses the data sparsity challenge by allowing generalization
across different queries and documents.  However, a linear model
in the features may not be sufficiently accurate to capture the
real CTR@1.  The bias terms thus provide a mechanism to correct
the residuals and to yield more accurate estimates.
Due to these bias terms, it may appear that Equation~\ref{eqn:model-hybrid} uses too many free parameters.  However, as will be cleared in the next subsection, we use regularization to control the magnitude of these terms, so the bias terms will be essentially zero except for popular query-document pairs.  Consequently, these terms can be used to yield a highly accurate CTR@1 estimate for popular $(q,u)$, while for unpopular $(q,u)$ (which suffer the data sparsity issue most) we essentially use the linear estimate $\pmb{\beta}^{\mt}\vecb{x}_{qu}$.  Such a dichotomy is done automatically within the regularization framework.

\subsection{Parameter Update Rule}\label{subsec:param-update}

This subsection addresses the problem of parameter updates for model (\ref{eqn:model-hybrid}).  We first describe how to fit the parameters if we are given a static set of data, then extend the update rule to the online case when data arrive sequentially, and finally discuss a few practical issues when deploying the update rules in large-scale ranking systems.

\subsubsection{Batch Parameter Fitting}

Suppose we are given a set $\Dset$ of $t$ data
in the form of $\{(q_i,u_i,c_i)\}_{i=1,2,\ldots,t}$, where $c_i\in\{0,1\}$ is the click feedback for $(q_i,u_i)$ provided by the $i$-th user.
Let $\Pset$ be the set of \emph{distinct} $(q,u)$ pairs observed in $\Dset$, and $N=|\Pset|$.
For brevity, denote the feature vector for the $(q_i,u_i)$ pair as $\vecb{x}_i$. 



A standard approach to learn the parameters in Equation~\ref{eqn:model-hybrid} for CTR@1 estimation is the ridge regression by using $\{c_i\}$'s as targets: we seek the optimal parameters that minimize a regularized square loss:
\begin{eqnarray}
f_t(\pmb\beta,\{b_{q,u}\}) &\defeq& \sum_{i=1}^t \left(c_i-\pmb\beta^\mt\vecb{x}_i-b_{q_iu_i}\right)^2 + \lambda_1\big\|\pmb\beta-\pmb\beta^{(0)}\big\|_2^2+\lambda_2\sum_{(q,u)\in\Pset}\big\|b_{qu}-b_{qu}^{(0)}\big\|_2^2, \label{eqn:rr}
\end{eqnarray}
where 
$\lambda_1$ and $\lambda_2$ are positive regularization parameters provided by uers, $\pmb\beta^{(0)}$ and $b_{qu}^{(0)}$ are the prior values, and $\|\cdot\|_2$ is the ordinary $\ell_2$-norm.  Here, regularization is applied to avoid over-fitting and to ensure numerical stability.



%


Since (\ref{eqn:rr}) is a least-squares problem with $d+N$ many parameters, one may think it is intractable to solve for the exact solution since the computation complexity is $O((d+N)^3)$ and $N$ is often very large.  Fortunately,
%
using matrix algebra,
we can derive a closed-form solution for the minimizer of (\ref{eqn:rr}), whose complexity is cubic in $d$ and only linear in $N$.

%
Specifically, we partition the index set $\{1,2,\ldots,t\}$
into $I_1 \cup I_2 \cup \cdots \cup I_N$, so that $I_j$ contains
indices in $\Dset$ that corresponds to the $j$-th distinct $(q,u)$ pair.
For every $j\in\{1,2,\ldots,N\}$, we define the following quantities,
\begin{eqnarray*}
a_j &\defeq& \lambda_2+ |I_j|, \\
\vecb{b}_j &\defeq& \sum_{i \in I_j} \vecb{x}_i, \\ 
d_j &\defeq& \lambda_2 b^{(0)}+\sum_{i \in I_j} c_i.
\end{eqnarray*}
In addition, we define
\begin{eqnarray*}
\vecb{A}_0 &\defeq& \lambda_1 \vecb{I} + \sum_{i=1}^t \vecb{x}_i\vecb{x}_i^\mt, \\
\vecb{d}_0 &\defeq& \lambda_1 \pmb\beta^{(0)} + \sum_{i=1}^t c_i\vecb{x}_i.
\end{eqnarray*}
%
%
Now the optimal solution to the least-squares problem must satisfy the first-order optimality condition:
\begin{eqnarray*}
\frac{\partial f_t(\pmb\beta,\{b_{q,u}\})}{\partial\pmb\beta} &=& \pmb{0}, \\ 
\frac{\partial f_t(\pmb\beta,\{b_{q,u}\})}{\partial b_j} &=& 0,
\end{eqnarray*}
for each $j\in\{1,\ldots,N\}$.  Solving this system of linear equations immediately gives the regularized least-squares solution:
\begin{eqnarray}
\pmb\beta^* &=& \left(\vecb{A}_0-\sum_{j=1}^N a_j^{-1}\vecb{b}_j\vecb{b}_j^\mt\right)^\mi\left(\vecb{d}_0-\sum_{j=1}^Na_j^\mi d_j\vecb{b}_j\right) \label{beta}\\
b_j^* &=& a_j^\mi \left(d_j - \vecb{b}_j^\mt\pmb\beta^*\right), \ \ \text{for each} \ \ j  \in \{1,2,\ldots,N\}.\label{bias}
\end{eqnarray}
In other words, the complexity of solving the least-squares problem now becomes $O(d^3+dN)$, a substantial improvement over the $O((d+N)^3)$ complexity of the naive approach.

\subsubsection{Online Parameter Updates}

More importantly, the formulas above suggest that we only need to maintain
a set of sufficient statistics ($\vecb{A}_0$, $\vecb{d}_0$, $a_j$, $\vecb{b}_j$, and $d_j$) to obtain the \emph{exact} solution when a new data is added to the set $\Dset$, without the need to re-computing all quantities.

When a new example $(q_{t+1},u_{t+1},c_{t+1})$ arrives, all these sufficient statistics can be updated efficiently in an incremental fashion.  In particular, let ${j_{t+1}}$ be the index of $(q_{t+1},u_{t+1})$ in $\Pset$, then $O(d^2)$ time is needed for the updates:
\begin{eqnarray*}
\vecb{A}_0 &\leftarrow& \vecb{A}_0 + \vecb{x}_{t+1}\vecb{x}_{t+1}^\mt \\
\vecb{d}_0 &\leftarrow& \vecb{d}_0 + c_{t+1}\vecb{x}_{t+1} \\
a_{{j_{t+1}}} &\leftarrow& a_{{j_{t+1}}} + 1 \\
\vecb{b}_{{j_{t+1}}} &\leftarrow& \vecb{b}_{{j_{t+1}}} + \vecb{x}_{t+1} \\
d_{{j_{t+1}}} &\leftarrow& d_{{j_{t+1}}} + c_{t+1}.
\end{eqnarray*}

With these updated sufficient statistics, we can now apply Equations~\ref{beta} and \ref{bias} to compute the exact solutions, which again requires $O(d^3+dN)$ time.  However, there are acceleration techniques that can reduce the complexity to $O(d^2)$ and even $O(d)$, as explained in the next subsection.

\subsubsection{Implementation Issues in Practice}

While the update rules derived above are reasonably efficient, we would still like greater acceleration for large $d$ and large $N$, so that the response time of the whole re-ranking system can be further reduced.

First of all, the most time-consuming part is the inversion of the matrix in Equation~\ref{beta}, which takes $O(d^3)$ time.  Fortunately, since every new data results in a rank-one update on the matrix, $\vecb{A}_0-\sum_{j=1}^N a_j^{-1}\vecb{b}_j\vecb{b}_j^\mt$, a straightforward variant of the famous Sherman-Morrison formula may be applied to reduce the complexity to $O(d^2)$.

Second, we may also ignore off-diagonal elements of the matrix, $\vecb{A}_0-\sum_{j=1}^N a_j^{-1}\vecb{b}_j\vecb{b}_j^\mt$, and so inversion can be done very efficiently in $O(d)$ time.  According to our experience (not reported in the present paper), this approximation is quite effective, yielding a good tradeoff between solution quality and time requirement.

Third, we note that it is unnecessary to update all $N$ bias terms $b^*_j$ every time a new example arrives.  In fact, these bias terms can be updated independently, provided that $\pmb{\beta}^*$ is given.  Therefore, we may delay their updates until the moment they are used.  Specifically, for a new example $(q_{t+1},u_{t+1},c_{t+1})$, we may only update $b^*_{{j_{t+1}}}$, where ${j_{t+1}}$ is the index of $(q_{t+1},u_{t+1})$ in $\Pset$.  This lazy-update trick completely removes the time dependency on $N$, a significant improvement when $N$ is large.


%

Finally, 
we note that it may even be impossible and unnecessary to explicitly maintain a bias term for every $(q,u)$ pair, since only a small fraction of them are popular queries and thus are expected to take advantage of those bias terms.
A few techniques such as the hashing trick~\cite{LanLiStr07} may be used to limit the effective number of bias terms.


\subsection{Variations of model}\label{subsec:variation}

Given the model form and update formula in Section \ref{subsec:param-ctr} and Section \ref{subsec:param-update}, there are a couple of choices to try for our online re-ranking function, which we describe below.

\noindent\textbf{Exploration and $\epsilon$-greedy: } In Section \ref{subsec:param-update}, we did not describe how the click feedbacks for the data $\mathcal{D}$ are collected. In order to explore rankings other than the output of our re-ranking model and collect balanced click feedback in our data set  $\Dset$, we use $\epsilon$-greedy strategy. 
The $\epsilon$-greedy is a simple strategy to handle the explore-exploit tradeoff described in Section \ref{subsec: notations}. It collects the feedback from the randomly permuted ranking with probability $\epsilon$ and from the re-ranked result by the function (\ref{eqn:model-hybrid}) with probability $1-\epsilon$. Thus, by controlling $\epsilon$, we can balance the exploration and exploitation for our online learning, and our exploration bucket data enables us to realize this strategy. More details on the methodology of using our exploration bucket data are described in the next section.

\noindent\textbf{Warm start: } When we are sequentially learning $(\pmb\beta, \{b_{qu}\})$ as described in Section \ref{subsec:param-update}, we need not learn them from scratch solely based on online learning (which is known as \emph{cold-start}), but learn a starting point from some already available click logs (i.e., \emph{warm-start}). The effectiveness of such warm-start models could be critical in terms of improving the performance and learning speed of our re-ranking function as presented in the next section.

\noindent\textbf{Using clicks on multiple positions:} In Section \ref{subsec:param-update}, we inherently assumed that the click feedback $\{c_i\}$'s are the ones received by the user when the document was displayed in the first position for the query, since we used them as a target for our CTR@1 function in (\ref{eqn:rr}). However, although our goal is maximizing CTR@1, we may not limit ourselves to use the click feedback only on position 1, that is, $c_{t,1}$ in the $r_t=\{c_{t,1}, \ldots, c_{t,s}\}$ defined in Section \ref{subsec: notations}, but use clicks on multiple positions for learning our re-ranking function. In that case, we can enlarge the data set $\Dset$ to $\{(q_i,u_{i,p},c_{i,p})\}_{i=1,2,\ldots,t}^{p=1,\ldots,s}$, and we introduce additional bias terms $\{b_p\}_{p=1}^s$ to correct the positional biases in the click feedback on position $p$. Then, we model CTR@$p$ as
\begin{eqnarray}
\text{CTR}@p(q,u) &=& \text{CTR}@1(q,u) + b_p \label{eqn:model-bias}
\end{eqnarray}
while modifying the loss function as
\begin{eqnarray}
\lefteqn{f_t(\pmb\beta,\{b_{qu}\},\{b_p\}) \,\,\defeq\,\, \sum_{i=1}^t \sum_{p=1}^s \left(c_{i,p}-\pmb\beta^\mt\vecb{x}_{i,p}-b_{q_{i,p}u_{i,p}}-b_p\right)^2} \nonumber \\
&&  + \lambda_1\big\|\pmb\beta-\pmb\beta^{(0)}\big\|_2^2 +\lambda_2\sum_{(q,u)\in\Pset_t}\big\|b_{qu}-b^{(0)}\big\|_2^2 + \lambda_3\sum_{p=2}^s\big\|b_p-b_p^{(0)}\big\|_2^2, \label{eqn:rr position}
\end{eqnarray}
where $\lambda_3$ and $b_p^{(0)}$ are regularization coefficient and prior for the positional bias terms $\{b_p\}$. Note that we set $b_p=0$ when $p=1$, and our re-ranking function is still CTR@1$(q,u)=  \pmb{\beta}^{T}\vecb{x}_{qu} + b_{qu}$ learned by minimizing (\ref{eqn:rr position}). In this way, we can utilize more click feedback than only using the clicks on position 1 to learn the re-ranking function (\ref{eqn:model-hybrid}). In Section \ref{sec: experimental results}, we will show how useful this approach is for building the warm start model described above. For the online updates, however, in order to control the number of experiments to compare, we remain to use only the clicks at the first positions and use the loss function and update formula in Section \ref{subsec:param-update} for all of the online schemes in our experiments.

\section{Experimental results}\label{sec: experimental results}

This section reports our experiments on various algorithms for
recency search re-ranking using the exploration bucket data described in
Section~\ref{subsec: bucket}.  Section \ref{subsec: evaluation
method} describes an unbiased offline evaluation method we will
adopt in the experiments.  Section~\ref{sec:methods} describes a
number of representative algorithms for comparison.  These
algorithms are selected to demonstrate benefits of various
algorithmic choices described in Section \ref{subsec:variation}.  Section~\ref{subsec:results} presents and
analyzes the experiment results in details.  Finally,
Section~\ref{sec:case-study} revisits the query examined in
Section~\ref{sec: motivation}, illustrating how our algorithm
adapts to user click feedback to re-rank the top documents and yield
better results.

\subsection{Unbiased Offline Evaluation}\label{subsec: evaluation method}

A tricky part of our problem is that, unlike in supervised learning,
it is hard to evaluate and compare performance of algorithms using
a \emph{static} set of log data.  The reason is that the click
feedback in the log depends on the ranking results that the user
observed when the log was collected; consequently, we do not know
what that user might have clicked if the algorithm we evaluate
ranked the results differently.  Fortunately, our exploration
bucket data can be used for reliable offline evaluation of
different algorithms, including both batch or online ones.

We follow the ``replaying'' evaluation method studied by
\cite{Li11Unbiased} for interactive applications like the
re-ranking problem considered here.  First, we hold out the
sessions for the latter three days in the exploration bucket data
and use it as a test set. The first three days' data may be used as a
training set for batch learning or warm start model described in Section \ref{subsec:variation}.  We then sort the test sessions
in the order of time stamps.
To evaluate an algorithm's CTR@1 on the test set, we maintain two
quantities, $C$ and $M$, which are interpreted as the number of
clicks at position 1 and number of search sessions, respectively.  Both $C$ and
$M$ are initialized to $0$.
\begin{enumerate}
\item We retrieve the $t$-th session in the test set, present the top $s$ documents together with their features to the re-ranking function.
\item The re-ranking algorithm then proposes to display one of the documents in the first position based on its re-ranking scores.  We call it a ``match'' if this proposed document is the same as the one displayed in the first position in the retrieved test session.
\item If a match happens, we reveal the user feedback $c_t$ (1 for click and 0 otherwise) to the algorithm, and perform the updates: $C \leftarrow C + c_{t}$ and $M \leftarrow M + 1$.
\item Otherwise, $c_t$ is not revealed, and the values of $C$ and $M$ are unchanged.  Effectively, this session is ignored.
\end{enumerate}
Finally, the overall CTR@1 of the algorithm in the evaluation
process above is $C/M$. 

For each session in our test set, the probability that a match
happens is $1/s$ for any ranking algorithm, since the top $s$
documents are randomly shuffled in our exploration bucket data.
Therefore, for a test set of $L$ sessions, $M$ equals $L/s$ on
average.  In our experiments, since $L$ is large, $M$ is almost
constant across different runs. The following key property
justifies the soundness of the evaluation method above: it can be
proved that the estimated CTR@1, $C/M$, of an online algorithm is
an unbiased estimate of its true CTR@1 \emph{as if we were able to
run it to serve live user traffic}~\cite[Theorem 2]{Li11Unbiased}.
Therefore, algorithms that have higher CTR@1 estimates using this
evaluation method will have higher CTR@1 in live buckets as well.
This important fact allows us to reliably compare and evaluate
various algorithms \emph{offline} without the costs and risks of
actually testing them with \emph{live} users.



\subsection{Models}\label{sec:methods}

There are various options to leverage user click feedback to
adjust a re-ranking function.  For instance, one may expect better
adaptation to user interest if a re-ranking system can adjust its
ranking function in real time based on user feedback; it may also
be interested in understanding how re-ranking performance is
affected by the CTR model, such as the ability to generalize (via
the linear features) and specialize (via the bias terms) in our
model (\ref{eqn:model-hybrid}).


Below, we describe a few representatives, chosen carefully to
demonstrate the benefits of various algorithmic choices.  The
methods are grouped into four categories.  

\begin{enumerate}

\item[1.] The first is a baseline
that is based entirely on editorial judgments and does not
leverage user clicks at all:


%

\begin{itemize}

\item \textbf{frmsc(baseline)}: We used the recency ranking function \cite{Anlei10} deployed in our search engine as a baseline.  This function was trained using time-varying recency features and recency demoted labels provided by human editors.  This method does not use click feedback.

\end{itemize}

\item[2.] The second category contains methods that learn CTR@1 from the first three days' 
training data and then do not online update in the test phase.
That is, the data $\mathcal{D}$ used in the update formula of
Section~\ref{subsec:param-update} only consists of the clicks in the first position in
the training set.  Such methods will be compared to their
online-learning counterpart.

\begin{itemize}

\item \textbf{batch(b)}: This is the linear model in (\ref{eqn:model-hybrid}) trained on the training set, and then deployed on the test set without any online updates. Note that there is no positional biases in the click feedback in the training set for this model due to the exploration bucket.

\item \textbf{batch(nb)}: This is the same as above, but does not use the bias terms in (\ref{eqn:model-hybrid}).  In other words, only a linear combination of features is used to compute a CTR@1 estimate.  This model is used to show the benefits of the bias terms.

\end{itemize}

\item[3.] The third category contains online learning methods in Section \ref{subsec:param-update} for
re-ranking with $\epsilon$-greedy strategy mentioned in Section \ref{subsec:variation}. We realize the $\epsilon$-greedy strategy in our online learning method by utilizing the exploration bucket data again. That is, while we use the exploration bucket data for an unbiased evaluation of performances of various schemes as in Section \ref{subsec: evaluation method}, we use the data once more to incrementally train the online schemes, as in Section \ref{subsec:param-update}. More concretely, at time $t$, the click feedback at the first position $c_{t,1}$ is revealed to the online schemes for the model updates, no matter whether there is a ``match'' or not for the schemes so that the re-ranking function can observe the feedbacks for all possible randomly served documents in the top position to correctly learn the re-ranking based on CTR@1. Note that this is effectively simulating the $\epsilon$-greedy strategy with $\epsilon=1$ and a separate deployment test bucket for evaluation. Also, a clear but subtle point is that we are revealing the click feedback after the data point was used for the `replay' evaluation so that we are not training and testing with the same data point. Moreover, in practice, we note there is usually a time delay between
delivery of the ranking result and the receipt of user feedback.
To make our evaluation and online learning process closer to reality,
we make the user feedback $c_{t,1}$ is not revealed to the re-ranking
algorithm immediately.  Rather, these signals are revealed every
five minutes (based on the time stamps of the test sessions) for our simulation.

Based on a few algorithmic choices, we tested following variations to see the effect of online schemes.


\begin{itemize}

\item \textbf{online(b)} and \textbf{online(nb)}: These are the online algorithms that optimize the parameters in (\ref{eqn:model-hybrid}) incrementally based on user click feedback, with and without bias terms, respectively.  Note that both algorithms learn the parameters \emph{from scratch}.

\item \textbf{online(b,ws)} and \textbf{online(nb,ws)}: These are the same as \textbf{online(b)} and \textbf{online(nb)} but use warm-start initialization of the parameters.  Specifically, we used the batch-learned parameters (in \textbf{batch(b)} and \textbf{batch(nb)}, respectively) as $\pmb\beta^{(0)}$ and $b_{qu}^{(0)}$ in (\ref{eqn:model-hybrid}).  We set $\lambda_1=\lambda_2=10$.  These methods thus combine the prior knowledge extracted from previous data with the ability to learn online.


\item \textbf{online(b,ws,w0)}: This method is similar to \textbf{online(b,ws)} except that the weight vector $\pmb{\beta}$ is fixed to the warm-start $\pmb{\beta}^{(0)}$ learned by \textbf{batch(b)}.  Thus, this model performs limited online updates and is useful to demonstrate the benefits of online update of $\pmb{\beta}$.

\item \textbf{counting}: Motivated by click models, this method maintains the ratio of cumulative clicks and
views for each document-query separately. This is an online
learning model, but does not utilize query--document features for
generalization. It may suffer the ``cold-start'' problem on the
tailed queries. Essentially, this scheme is equivalent to only maintaining the bias terms for observed document-query pairs. 

\end{itemize}

\item[4.] Batch learning of (warm-start) parameters in the previous two
categories is trained on the first three days of exploration bucket data. However, although the exploration bucket gives the unbiased CTR@1 of the documents as in Section \ref{sec: motivation}, in practice, it is not realistic to always require such expensive data in order to build the batch model as a starting point for online models. Therefore, we use the \emph{controlled log} from
non-exploration buckets (e.g., production bucket) for the same period of time to build the batch models and compare them with the models built from the exploration bucket. Note that the controlled log is very cheap to attain, but have large positional biases in clicks, so it is not clear to see how well the batch models trained on the controlled logs would perform. We include following three variations regarding the batch model with the controlled log, which contained $485,135$ sessions from production logs that overlaps with the first three days of exploration bucket.
 \footnote{Note that the online methods described above
may also be combined with warm-start models learned from control
logs.  We do not include them for
comparison in the paper to make the presentation of the results simple.}.

%

\begin{itemize}

\item \textbf{batch(control@1)} This method learns the model in (\ref{eqn:model-hybrid}) using the clicks at the first positions in the controlled log.

\item \textbf{batch(control@4,np)} This method learns the model with the loss function in (\ref{eqn:rr position}) and using top four positions' clicks in the controlled log. However,  it ignores the position biases, $\{b_p\}$'s are all set to zero, so data from all four positions are not distinguished.

\item \textbf{batch(control@4)} This method improves on \textbf{batch(control@4,np)} by considering position biases and including nonzero $\{b_p\}$'s as described in Section \ref{subsec:variation}.

\end{itemize}

\end{enumerate}

\subsection{Experimental Results}\label{subsec:results}


We ran the algorithms described in the previous subsection, whose overall nCTR@1 results are summarized in Table~\ref{tab:dep-ctr}.  The lifts over the baseline's nCTR@1 are also computed.

\begin{table}[h]
\caption{Overall cumulative nCTR@1 on the test set.} \label{tab:dep-ctr}
\centering
\begin{tabular}{l|c|c}
\hline
algorithms & nCTR@1 & lift over \textbf{frmsc}\\
\hline\hline
\textbf{frmsc(baseline)} & $0.770$ & $0\%$\\
\hline
\textbf{batch(b)} & $0.877$ & $13.90\%$ \\
\textbf{batch(nb)} & $0.849$ & $10.26\%$ \\
\hline
\textbf{online(b)} & $0.875$ & $13.64\%$ \\
\textbf{online(nb)} & $0.839$ & $\phantom{0}8.96\%$ \\
\textbf{online(b,ws)} & $\mathbf{0.901}$ & $\mathbf{17.01\%}$ \\
\textbf{online(nb,ws)} & $0.851$ & $10.52\%$ \\
\textbf{online(b,ws,w0)} & $0.891$ & $15.71\%$ \\
\textbf{counting} & $0.872$ & $13.25\%$ \\
\hline
\textbf{batch(control@1)} & $0.883$ & $14.68\%$ \\
\textbf{batch(control@4,np)} & $0.856$ & $11.17\%$ \\
\textbf{batch(control@4)} & $0.885$ & $14.94\%$ \\
\hline
\end{tabular}
\end{table}

\begin{figure}[t]
\centering
\includegraphics[width=0.8\textwidth]{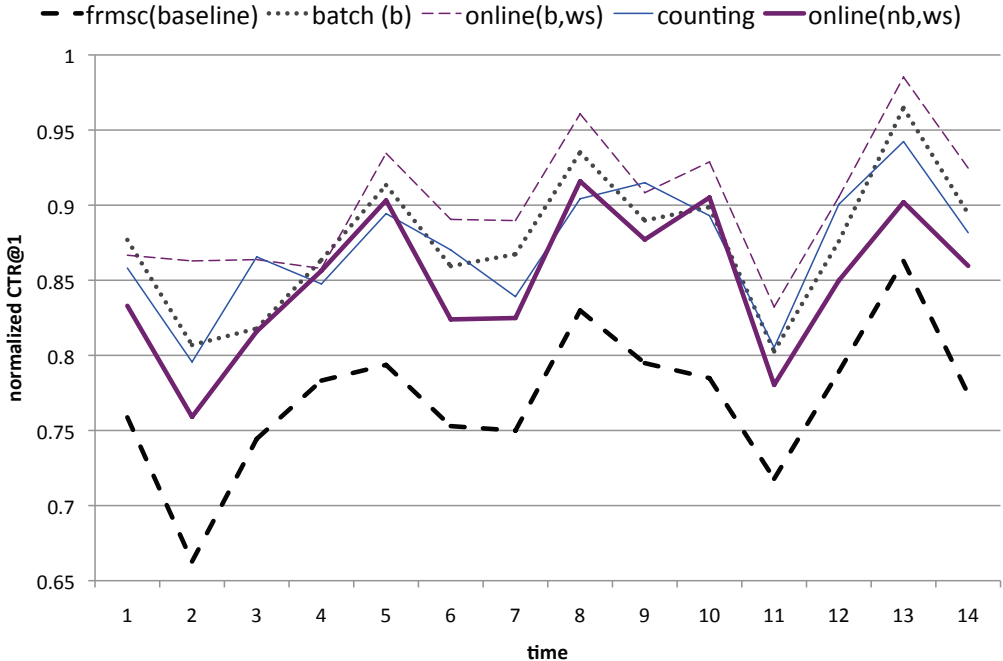}
\caption{Instantaneous (normalized) CTR@1 of the four models on
the test data. Each point was measured for an approximately
six-hour period.} \label{fig:inst-ctr}\end{figure}

To visualize how instantaneous nCTR@1 evolves over time, we also
computed aggregated clicks in every 6-hour period for the
algorithms.  Only five algorithms are included in
Figure~\ref{fig:inst-ctr} to ensure legibility.

\subsubsection{Results Analysis}

A number of important observations are in order based on the
results reported in Table~\ref{tab:dep-ctr} and
Figure~\ref{fig:inst-ctr}.

First, the advantage of leveraging user click feedback is obvious
from the lift of \emph{all} algorithms over the baseline that does
not use click feedback at all.  The click lift, which ranges from
$8.96\%$ to $17.01\%$, is statistically significant given the size
of our data set.  Even the batch-learning methods that do not
perform online updates are quite strong, capable of achieving at
least $10\%$ lift.

Second, we can see the additional benefits brought by the bias
terms in both batch and online algorithms.  It is true in all
cases that an algorithm is better than its counterpart without
bias terms.  In the case between \textbf{online(b,ws)} and
\textbf{online(nb,ws)}, the bias terms account for about $6.49\%$
lift.

Third, while batch algorithms have quite strong lift, greater
lifts are achieved by algorithms that adjust their re-ranking
functions \emph{online}.  This benefit is expected since the
online algorithms are able to extract more information from online
click feedback, in addition to the batch-learned models.    In
addition, it is worth pointing out a practically important fact
that it is \emph{compatible} to use batch-learned models as
warm-start models for online methods.  Of all the algorithms in
Table~\ref{tab:dep-ctr}, the greatest lift is achieved by
\textbf{online(b,ws)}---the online method that uses batch-learned
warm-start models and bias terms.  Even for the coefficient vector
$\pmb{\beta}$ that is shared by all query-document pairs, updating
it online is still helpful, which is justified by the gap between
\textbf{online(b,ws)} and \textbf{online(b,ws,w0)}.  A larger gap
is possible if the time span of our test data is larger.

Fourth, we examine the role of generalization in our CTR@1 model.
As discussed earlier, the bias terms are essentially zero except
for popular query-document pairs, due to the regularization we
used in the optimization step.  Therefore, the linear part in
(\ref{eqn:model-hybrid}) determines the CTR@1 estimates of tail
queries for which we observed one or few sessions.  The result in
Table~\ref{tab:dep-ctr} confirms our conjecture: the method
\textbf{counting} which uses the bias terms alone for re-ranking
yielded a lower click lift than \textbf{online(b,ws,w0)} or
\textbf{online(b,ws)}. The reason is that the CTR model in
\textbf{counting} can't make good prediction in ``cold-start''
situation and so little lift was achieved on the tailed queries.

Finally, our results suggest it is possible to use control log to
build a competitive warm-start model. It should be noted that the
control log has much more data than the exploration data, thus the
strong performance. However, position bias has to be considered
when clicks from multiple positions are used, as demonstrated by
the gap between \textbf{batch(control@4)} and
\textbf{batch(control@4,np)}. We believe the performance of
learning from controlled logs could be further improved by
advanced click models. We plan to investigate this direction in
future work.

\subsubsection{Lift Distribution}

\begin{figure}[t]
\centering
\includegraphics[width=0.75\textwidth]{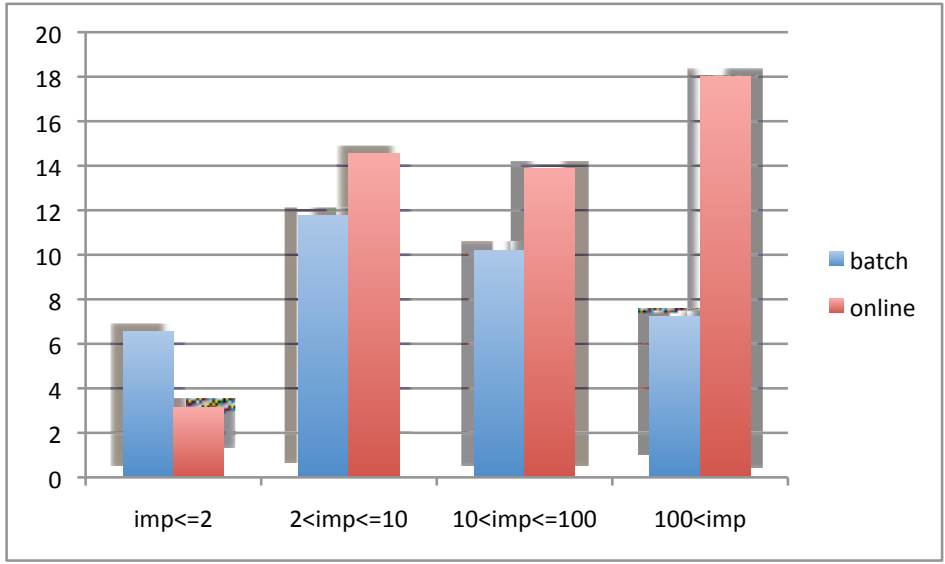}
\caption{Relative CTR@1 lift in percentage ($\%$) over the
baseline model (frmsc) for queries with different
impressions.}\label{fig:query_impression}
\end{figure}

\begin{figure}[h]
\centering
\includegraphics[width=0.75\textwidth]{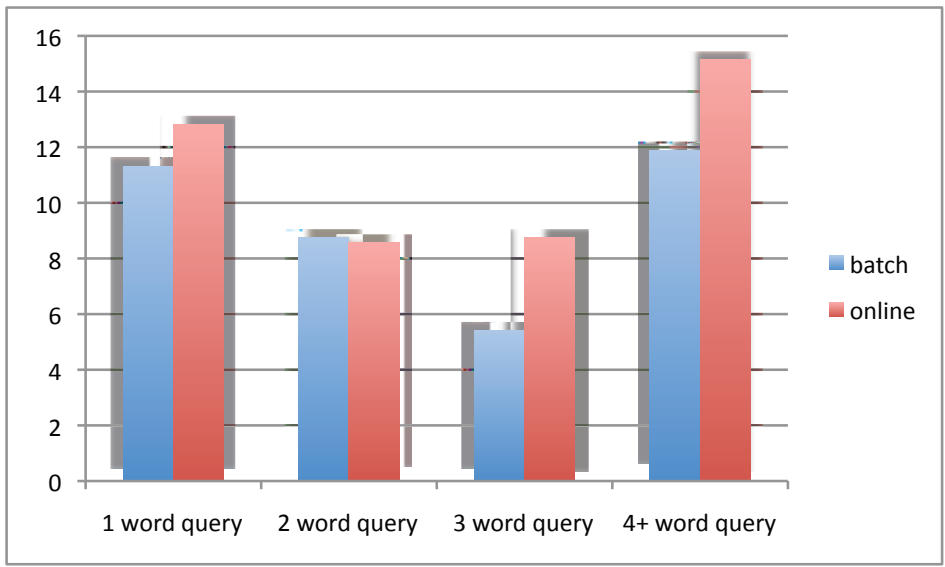}
\caption{Relative CTR@1 lift in percentage ($\%$) over the
baseline model (frmsc) for queries with different
lengths.}\label{fig:query_length}
\end{figure}

We now examine lift distribution over queries with different
impression and lengths, respectively.  Results of two
representatives, \textbf{batch(b)} and \textbf{online(b,ws)}, are
reported.

Figure~\ref{fig:query_impression} presents CTR@1 lifts over
\textbf{frmsc} aggregated over queries with different impressions.
We notice that the \textbf{online(b,ws)} model is doing very well
on popular recency queries, whereas the \textbf{batch(b)} model
gives more lift on queries with less than 2 impressions. For
queries with very limited impressions, e.g. less than 2, the
\textbf{online(b,ws)} model cannot gain much advantage over the
\textbf{batch(b)} model.
This problem might be mitigated when applying the
\textbf{online(b,ws)} model to larger traffic. This observation
also suggests a practical solution, i.e. employing
\textbf{batch(b)} models for queries with scarce impressions while
using \textbf{online(b,ws)} models for popular recency queries only. It is
expected that the \textbf{online(b,ws)} model achieves much more
lift on popular recency queries, since the \textbf{batch(b)} model
cannot specialize well on such cases despite the relatively large number of impressions. For example, the query
``\emph{giant squids in California}'' was a popular recency query.
A \textbf{batch(b)} model well trained on historical click events
still fails to foresee the popularity and high relevance of the
youtube video, whereas the \textbf{online(b,ws)} model does so
correctly by adapting to users' click feedback; see
Section~\ref{sec:case-study} for more details of this example.

Figure \ref{fig:query_length} presented CTR@1 lifts over
\textbf{frmsc} for queries with different lengths. Except for a
tie in two-word queries, the \textbf{online(b,ws)} model
consistently outperforms the \textbf{batch(b)} model.  The results
suggest the online re-ranking method is robust to queries of
various lengths.

\subsubsection{Comparison of Other Click Metrics}\label{subsec: click metric}

\begin{figure}[t]
\centering
\includegraphics[width=0.75\textwidth]{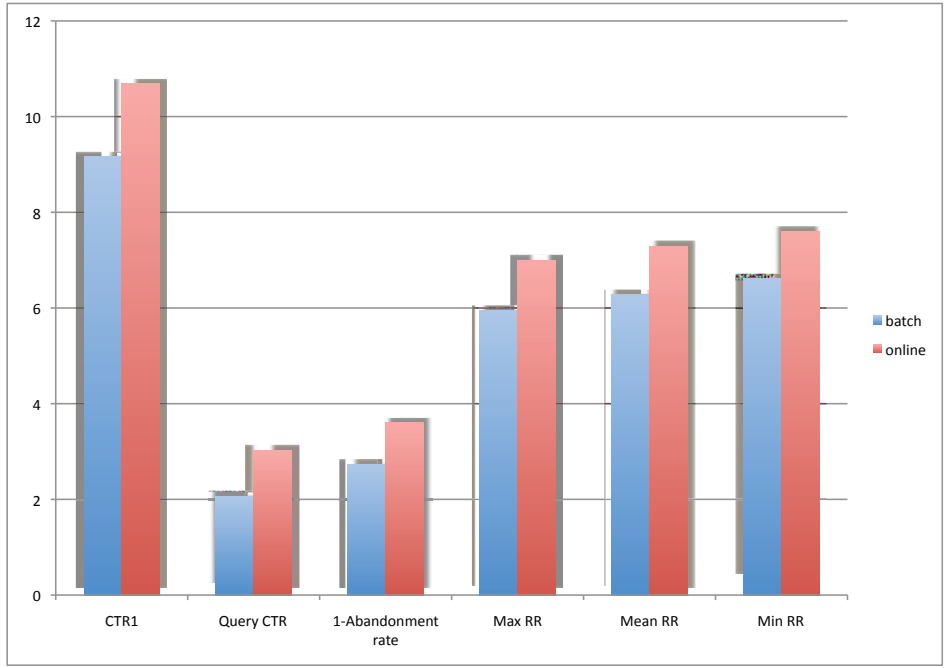}
\caption{Relative lift in percentage ($\%$) over the baseline
model (frmsc) on various click metrics.}\label{fig:click_metric}
\end{figure}

\begin{figure}[h!]
\centering 
\subfigure[Zoom-in plot]{\label{start fv}
\includegraphics[width=0.7\textwidth]{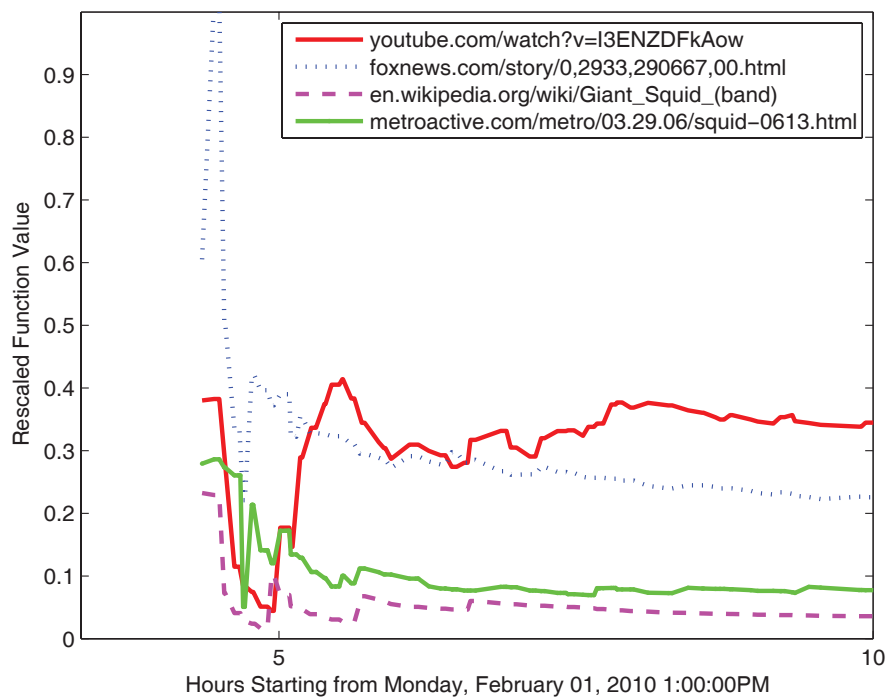}}
\subfigure[Whole period]{\label{entire fv}
\includegraphics[width=0.7\textwidth]{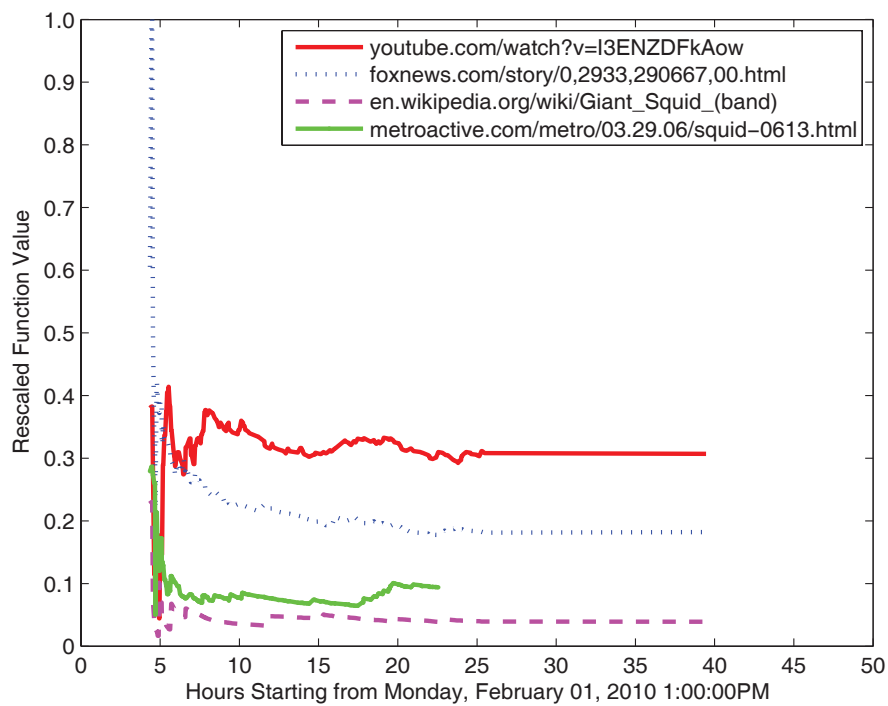}}
\caption{Function values in the online model on
the 4 URLs associated with the recency query ``\emph{giant squid
in California}''. Note that the URL of metroactive.com was
replaced by another URL (not shown here) after the 23rd
hour.}\label{fig:squidfunc}
\end{figure}

Although we have focused on training and evaluating our online
re-ranking function based on CTR@1, we also compare our results
with following other click metrics for ranking proposed in ~\cite{RadKurJoa08}:
\begin{itemize}
  \item \textbf{Query CTR} is the average number of clicks for each query
  \item \textbf{$1-$Abandonment Rate} is the probability of a session receiving a click
  \item \textbf{Max RR (Reciprocal Rank)} is the reciprocal rank of the highest ranked result clicked on
  \item \textbf{Mean RR (Reciprocal Rank)} is the average of clicked documents' reciprocal ranks
  \item \textbf{Min RR (Reciprocal Rank)} is the reciprocal rank of the lowest ranked results clicked on. 
 \end{itemize} 
  Thus, for all metrics, higher values are assumed to indicate better ranking qualities.

%





In Figure \ref{fig:click_metric}, both \textbf{batch} and
\textbf{online} models show significant lifts over the
\textbf{frmsc} baseline on all click metrics listed above.
Moreover, the \textbf{online(b,ws)} method consistently gives
about 2\% more lift than \textbf{batch(b)}. Thus, although our
algorithms focus on maximizing CTR@1, it also gives simultaneous
lifts on other click metrics as well. This fact shows the easily
measurable CTR@1 is a good surrogate to optimize, and also
justifies our choice of using click at the top position as user
feedback.

\noindent\emph{Ramark:} We have done similar experiments on the non-recency queries as well, but our online re-ranking scheme did not show too much gain for those queries as in the recency queries, of which results are omitted here. 
The absence of improvements for non-recency queries is expected:
the relevance of documents with respect to such queries does not change dramatically over time, so the re-ranking based on the click feedback may not be too different from the original ranking.

\subsection{Case Study} \label{sec:case-study}

We now revisit our example query ``\emph{giant squid in
California}'' in Section \ref{sec: motivation} to illustrate how
our online re-ranking function can adapt to the click feedback
quickly and track the best re-ranking. Coincidentally, the query
happened to only appear in our test set. We ran our online model,
\textbf{online(b,ws)}, on the test set, and recorded the function
values of the 4 URLs. Figure \ref{start fv} shows the function
values of 4 URLs for the first 10 hours, and Figure \ref{entire
fv} presents the entire temporal curves for those function values
in the lifetime of ``\emph{giant squid in California}''.

Since the initial re-ranking function \textbf{online(b,ws)} is
(almost) identical to the fixed one of \textbf{batch(b)}, we can
see from Figure \ref{start fv} that when the query appears around
the 5th hour, \textbf{batch(b)} orders the four URLs as
\begin{enumerate}
\item \verb"foxnews.com/story/0,2933,290667,00.html"
\item \verb"youtube.com/watch?v=I3ENZDFkAow"
\item \verb"metroactive.com/metro/03.29.06/squid-0613.html"
\item \verb"en.wikipedia.org/wiki/Giant_Squid_(band)".
\end{enumerate}
Note that this ranking is different from the \textbf{frmsc}
ranking presented in Section \ref{sec: motivation}. That is,
although the \textbf{batch(b)} has not observed the query
``\emph{giant squid in California}'' in its training set sessions,
from the sessions of other queries in the training set, it was
able to predict based on the query--document features that the
``video'' page will attract many clicks for the query and improve
the original ranking. Nonetheless, we see that it still fails to
accurately predict the users' click behaviors.

On the other hand, given the users' click patterns in Figure
\ref{fig:squidctr}, the \textbf{online(b,ws)} promptly learns from
them and put the ``video'' content with the highest CTR to the top
rank within an hour. The ranking was then maintained for the rest
of the time. Then, after the 25th hours, when the impression of
the query quickly decreased toward 0 as shown in Figure
\ref{fig:squid}, the function values of \textbf{online(b,ws)} were kept intact as can be seen in Figure \ref{entire fv}.

Thus, this example indeed illustrates how re-ranking algorithms
may benefit from user click feedback to improve ranking results.
Furthermore, by real-time adaption \textbf{online(b,ws)} can
quickly learn from users' click patterns and outperforms not only
the editorial-based batch recency ranking, \textbf{frmsc}, but
also the click-based batch re-ranking, \textbf{batch(b)}, for
recency queries.

\section{Related work}\label{sec: related work}

As mentioned in Section~\ref{sec: introduction}, the machine-learned
ranking framework has been extended to the recency search problem.
Previous work~\cite{Anlei10} introduced query classifiers
to detect time-sensitive queries, implemented time-varying
features that reflect document freshness, and recency demoted
labels provided by human editors.  More recently, improvements
are made by introducing additional click
related time-varying features~\cite{yoshi10,AnleiTwitter10}.
Algorithms studied in our work differ from them in two ways: (i) we use user click as training labels to gain further improvement; (ii) instead of fixing the ranking function and adjusting time-varying features, our features need not change over time, but the parameters of the ranking function can be automatically adjusted in real time based on user clicks.

Another related work \cite{Diaz09} considered an online
algorithm of slotting news direct display modules for recency
queries in the search results page but did not consider
re-rankings of documents.

Using users' click feedback to improve ranking quality of a search
engine has been extensively studied before.
User behavior models (\textit{e.g.}, 
\cite{Dupret2010WSDM,Chapelle2009,clickchain09}) are developed based on click log data, whose outputs were then used as features for batch-training a ranking function.  However, it is not easy to reflect temporal variations of document relevance in these works since the features were often computed in an average sense.

A few other works also used click data to directly modify their ranking based on the inference on the users' relative preferences on rankings, but their settings or focuses are different: the method of \cite{shihao2009} remained in the batch-learning mode and did not consider the temporal dimension of the click data; \cite{RadJoa07}
was similar in spirit to ours but did not consider strategies that generalize to tail queries, and their results were based on \emph{simulated} user clicks rather than \emph{real} ones; finally, the dueling bandit approach~\cite{YueJoa09} required a special functionality of the retrieval system to interleave two different ranking results.


%

Taking temporal variation of relevance into account to produce
better rankings has also been considered in the past. \cite{ElDu2010} considered the temporal variations of document content and applied that knowledge to improve search ranking, but did not utilize click feedback to directly refine the ranking. \cite{Kor09} devised a scheme to capture temporal dynamics of user ratings on items in a
collaborative filtering problem, but focused rather on long-term
dynamics and did not consider the cold-start problem, which is critical to our recency ranking application give the large volume of of tail queries.
Personalized article recommendation on web portals
is another closely related problem.
While models in earlier works (\textit{e.g.}, \cite{icdm09}) did not generalize,
there have been efforts on generalization more recently, such as the LinUCB algorithm~\cite{LiChuLanSch10} that uses a similar linear model as ours, and the warm-start solution by \cite{AgaCheEla10}.
However, both work remain
to maintain models for a small number of articles/items, and so have not demonstrated capacities of learning with an almost infinitely large
content pool, as in the space of query-document pairs in search
domain. Another difference is that their model was more
item-specific, whereas our model consists of both global
model that applies to all queries and documents and specific bias
term for each query-document pair. As a similar, independent thread of work, \cite{SteHerGra09} also considered the large-scale personalized recommendation problem, but imposed a Bayesian framework, which is different from our work.

\section{Conclusions} \label{sec: conclusion}

In this work, we investigated various learning algorithms to
re-ranking recency search results based on real-time user
feedback. Our contributions are three-fold. First, our evaluation
method is novel for web search---a random exploration bucket was
used to collect user feedback, which not only removed positional
bias but also allowed one to reliably evaluate online learning
algorithms \emph{offline}. Second, we proposed a re-ranking
approach to improve current search results for recency queries,
and carried out extensive empirical results for a dozen of
variants. Third, we demonstrated the need for using online
learning as a flexible machine learning paradigm to adapt a
ranking system to time-varying document relevance.

In future work, we would like to explore other options for correcting
position biases and using clicks on multiple positions, e.g.,
using multiplicative bias correction terms or using user click
models (e.g., \cite{Chapelle2009}), so that we can effectively
increase the size of training data and thus may result in faster
learning speed in practice. In this work we focused on ranking documents based on
individual document's CTR estimate.  It is also much more challenging
to design algorithms for the best permutation of a set of
documents, in which interactions between documents can be taken
into account.

\bibliographystyle{plain}
\bibliography{cumrel,bibfile}









\end{document}

%% file: math.tex
\newcommand{\mi}{{-1}}  
\newcommand{\mt}{{\top}}  

\newcommand{\vecb}[1]{{\mathbf{#1}}}

\newcommand{\defeq}{{\stackrel{\mathrm{def}}{=}}}